\documentclass[reprint,twocolumn,prl,showpacs,amsmath,amssymb,aps]{revtex4-1}

\usepackage{natbib}	
\usepackage{graphicx,color,rotating}
\usepackage[latin1]{inputenc}
\usepackage{textcomp}
\usepackage{dcolumn}
\usepackage{soul}
\usepackage{bm}     
\usepackage{upgreek}
\usepackage{subdepth}  			
\usepackage[latin1]{inputenc}

\usepackage{hyperref}						
\usepackage{xcolor}						
\hypersetup{							
    colorlinks,
    linkcolor={blue!80!blue},
    citecolor={blue!80!blue},
    urlcolor={blue!80!blue}
}
\newcommand{\mr}[1]{\ensuremath{\mathrm{#1}}}
\renewcommand{\vec}[1]{\bm{#1}}
\newcommand{\ee}{\mathrm{e}}

\newcommand{\ii}{\mathrm{i}}

\newcommand{\abs}[1]{\big|{#1}\big|}

\newcommand{\pscale}{p_1^0}
\newcommand{\vscale}{v_2^0}
\newcommand{\Fscale}{F_\mr{rad}^0}

\newcommand{\pp}{\partial}

\newcommand{\nablabf}{\boldsymbol{\nabla}}

\newcommand{\Lapl}{\nabla^2}

\newcommand{\eg}{\textit{e.g.}}

\newcommand{\etal}{\textit{et~al.}}

\newcommand{\qmarks}[1]{``{#1}"}

\newcommand*{\plimsoll}{{\ensuremath{-\kern-4pt{\ominus}\kern-4pt-}}}
\newcommand{\scap}{\!\cdot\!}

\newcommand{\cfl}{c_\mr{fl}}

\newcommand{\dd}{\mr{d}}

\newcommand{\eey}{\vec{e}_y}

\newcommand{\FFF}{\vec{F}}

\newcommand{\FFFrad}{\vec{F}_\mathrm{rad}}
\newcommand{\Frad}{F_{\mathrm{rad}}}

\newcommand{\kc}{k_\mathrm{c}}

\newcommand{\ks}{k_\mathrm{s}}

\newcommand{\nnn}{\vec{n}}

\newcommand{\rrr}{\vec{r}}

\newcommand{\uuu}{\vec{u}}

\newcommand{\vvv}{\vec{v}}

\newcommand{\zerovec}{\boldsymbol{0}}

\newcommand{\calC}{\mathcal{C}}

\newcommand{\calV}{\mathcal{V}}

\newcommand{\Eac}{E_\mathrm{ac}}

\newcommand{\Ccost}{\calC}

\newcommand{\FFFdrag}{\FFF_{\mathrm{drag}_{}}}

\newcommand{\kapfl}{\kappa_\mathrm{fl}}

\newcommand{\etafl}{\eta_\mr{fl}}

\newcommand{\etaBfl}{\eta^\mathrm{b}_\mathrm{fl}}

\newcommand{\Gamfl}{\Gamma_\mathrm{fl}}

\newcommand{\pfl}{p_\mathrm{fl}}

\newcommand{\rhofl}{\rho_\mathrm{fl}}

\newcommand{\kO}{k_{0}}

\newcommand{\SIMHz}{\textrm{MHz}}

\newcommand{\SImm}{\textrm{mm}}
\newcommand{\SImum}{\textrm{\textmu{}m}}

\newcommand{\SIPa}{\textrm{Pa}}

\newcommand{\SIpN}{\textrm{pN}}
\newcommand{\SIMPa}{\textrm{MPa}}

\newcommand{\SIs}{\textrm{s}}

\newcommand{\nn}{\nonumber}
\newcommand{\bsub}{\begin{subequations}}
\newcommand{\esub}{\end{subequations}}

\newcommand{\eqlab}[1]{\label{eq:#1}}
\renewcommand{\eqref}[1]{Eq.~(\ref{eq:#1})}
\newcommand{\eqnoref}[1]{(\ref{eq:#1})}

\newcommand{\figref}[1]{Fig.~\ref{fig:#1}}

\newcommand{\figlab}[1]{\label{fig:#1}}

\newcommand{\grad}{\boldsymbol{\nabla}}
\newcommand{\pargrad}{\boldsymbol{\nabla}_\parallel}

\renewcommand{\div}{\nablabf\!\cdot}

\newcommand{\lap}{\nabla^2}

\renewcommand{\Re}{\mathrm{Re}}

\begin{document}

\title{Suppression of acoustic streaming in shape-optimized channels}

\author{Jacob S. Bach}
\email{jasoba@fysik.dtu.dk}
\affiliation{Department of Physics, Technical University of Denmark, DTU Physics Building 309, DK-2800 Kongens Lyngby, Denmark}

\author{Henrik Bruus}
\email{bruus@fysik.dtu.dk}
\affiliation{Department of Physics, Technical University of Denmark, DTU Physics Building 309, DK-2800 Kongens Lyngby, Denmark}

\date{25 February 2020}

\begin{abstract}
Acoustic streaming is an ubiquitous phenomenon resulting from time-averaged nonlinear dynamics in oscillating fluids. In this theoretical study, we show that acoustic streaming can be suppressed by two orders of magnitude in major regions of a fluid by optimizing the shape of its confining walls. Remarkably, the acoustic pressure is not suppressed in this shape-optimized cavity, and neither is the acoustic radiation force on suspended particles. This basic insight may lead to applications, such as acoustophoretic handling of nm-sized particles, which is otherwise impaired by acoustic~streaming.
\end{abstract}


\maketitle


When a fluid executes oscillatory motion due to an imposed acoustic field or a vibrating boundary, the inherent fluid-dynamical nonlinearities spawn a steady flow adding to the oscillatory motion. This phenomenon, called acoustic streaming, has a rich, 200 year old history. Early observations by Ørsted (1809) and Savart (1827) of the
difference in the motion of coarse and fine grained powders over vibrating Chladni plates, were in 1831 conclusively attributed to acoustic streaming in the air by Faraday in his seminal experiments on Chladni plates placed in a partial vacuum~\cite{Faraday1831}. In 1876, Dvo\v{r}\'{a}k observed acoustic streaming caused by standing sound waves in a Kundt's tube~\cite{Dvorak1876}. A theoretical explanation of this boundary-induced streaming in various geometries was provided in 1884 by Lord Rayleigh in terms of an oscillatory boundary layer flow, which by time-averaging induces a steady slip velocity near the boundary that drives the steady streaming~\cite{LordRayleigh1884}. A further experimental and theoretical analysis of streaming invoking Prandtl boundary layers was presented by Schlichting in 1932~\cite{Schlichting1932}, who identified counter-rotating vortices inside the thin viscous boundary layer near the wall co-existing with the rotating vortices outside the boundary layer. Rayleigh's slip-velocity formalism was
later generalized to curved surfaces moving in the normal direction~\cite{Nyborg1958, Wang1986},
to flat surfaces moving in arbitrary directions~\cite{Vanneste2011}, and to curved surfaces with arbitrary velocity~\cite{Bach2018}. Eckart found in 1948 that acoustic streaming also can be induced by attenuation of sound in the bulk~\cite{Eckart1948}. This effect is mainly considered important for systems much larger than the acoustic wavelength~\cite{Westervelt1953, Riaud2017}, but as was pointed out recently,  it can also be significant on the length scale of a single wave length for rotating acoustic fields~\cite{Bach2019}.

Acoustic streaming is a truly ubiquitous phenomenon that has been observed not only in Newtonian fluids, but also in superfluid helium~\cite{Rooney1982} and non-Newtonian viscoelastic liquids~\cite{Vishwanathan2019}. It has  found many applications within a wide range of topics such as thermoacoustic engines~\cite{Ramadan2018}, enhancement of electrodedeposition~\cite{Kaufmann2009}, mixing in microfluidics~\cite{Sritharan2006}, biofouling removal~\cite{Sankaranarayanan2008}, and lysing of vesicles~\cite{Marmottant2003}. Given its widespread appearance, a fundamental question naturally arises: is it possible to suppress acoustic streaming? Recently, Karlsen \etal\ showed experimentally and
theoretically that for inhomogeneous fluids inside a microchannel, the acoustic streaming can be suppressed in the bulk of the fluid as long as a density gradient is present there~\cite{Karlsen2018}, an effect caused by the acoustic body force~\cite{Karlsen2016}.

But what about homogeneous fluids? In this work, using the same experimentally-validated numerical modeling as in Refs.~\cite{Karlsen2016, Karlsen2018}, we demonstrate that for homogeneous fluids confined in cavities or channels, the acoustic streaming can be suppressed by more than two orders of magnitude in large parts of the bulk by optimizing the shape of the confinement. This discovery not only provides physical insight into a time-honored fundamental phenomenon in fluid dynamics, but it is also of considerable practical interest in the field of microscale acoustofluidics, where ultrasound fields routinely are used to handle suspended microparticles. Such a particle of radius $a$ is affected by two forces: the acoustic radiation force that scales with $a^3$ and tends to focus particles at the acoustic nodal planes; and the streaming-induced drag force that scales with $a$ and by virtue of the streaming vortices tends to defocus particles. Consequently, there exists a lower limit of $a$ that allows for controlled handling by the focusing radiation force, and it has been shown to be $a_\mr{min} \approx 1~\SImum$ for dilute aqueous particle solutions~\cite{Muller2012,Barnkob2012a}. A suppression of the acoustic streaming would enable a desirable controlled handling of nanoparticles, such as bacteria, viruses and exosomes.

\begin{figure*}[t]
\includegraphics[width=\textwidth]{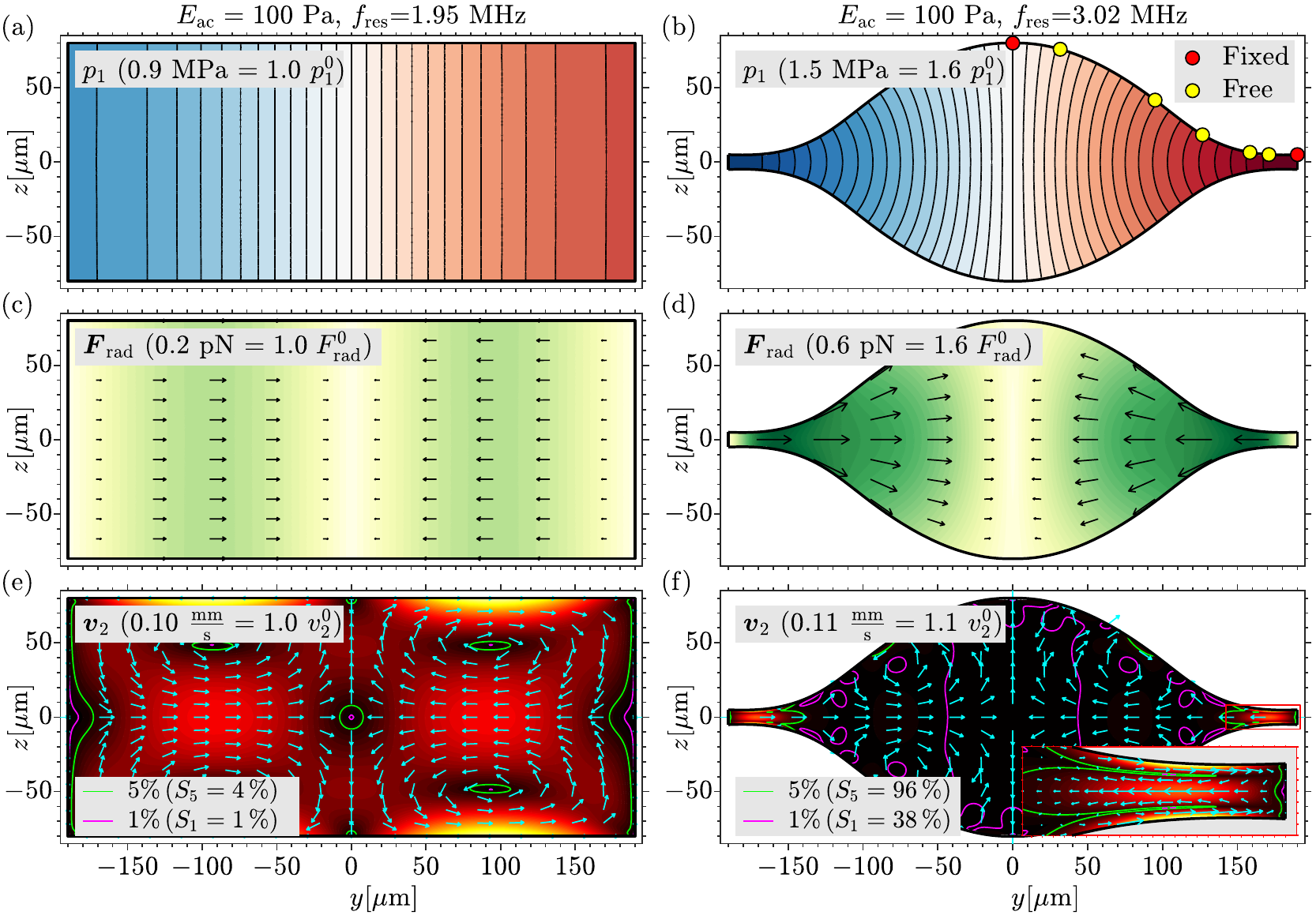}
\caption{\figlab{rect_vs_shape} Simulation results for MHz-acoustics at $\Eac=100~\SIPa$ in straight microchannels with a rectangular and a shape-optimized cross section. (a)-(b) The acoustic pressure $p_1$ from $-1.5~\SIMPa$ (light blue) to $+1.5~\SIMPa$ (dark red). The shape in (b) is defined by a spline interpolation between the colored points, where the $z$-coordinate of the yellow points are free in the optimization, and the red points are fixed. (c)-(d) The radiation force $\FFFrad$ (black arrows) on 1-$\SImum$-diameter polystyrene particles from $0$ (white) to $0.6~\SIpN$ (dark green). (e)-(f) The acoustic streaming $\vvv_2$ (cyan unit arrows) from $0$ (black) to $0.11~\frac{\SImm}{\SIs}$ (light yellow). The two contours mark $1~\%$ (magenta) and $5~\%$ (green) of the characteristic streaming speed $\vscale$ from \eqref{scaling}.}
\end{figure*}

\textit{Modeling the acoustofluidic fields.}---To optimize the shape,
efficient and fast computation of the acoustofluidic fields is required. For that, we use the  method described in Refs.~\cite{Bach2018, Skov2019}, where the thin viscous boundary layer is taken into account analytically  and therefore needs not to be resolved numerically. We consider a domain $\Omega$ with hard boundary walls, see \figref{rect_vs_shape}, containing a homogeneous and quiescent fluid of dynamic viscosity $\etafl$, bulk viscosity $\etaBfl$, density $\rhofl$, and sound speed $\cfl$ at pressure $\pfl$. An acoustic field is created by letting the boundary oscillate harmonically with the angular frequency $\omega$ around its equilibrium position $\pp\Omega$
with a prescribed displacement $\uuu^\mr{bdr}(\rrr,t)$ expressed as the real part  of the complex amplitude $\uuu^\mr{bdr}_1(\rrr)$,
\begin{equation} \eqlab{uuu}
\uuu^\mr{bdr}(\rrr,t)= \Re\big[\uuu^\mr{bdr}_1(\rrr)\: \ee^{-\ii\omega t}\big],
\end{equation}
where $\ii=\sqrt{-1}$. The resulting pressure $p$ is written as a perturbation series,
 \begin{equation} \eqlab{perturbation}
 p(\rrr,t)=\pfl +\Re\big[p_1(\rrr) \ee^{-\ii\omega t}\big]+p_2(\rrr),
 \end{equation}
and likewise for the density $\rho$ and the fluid velocity $\vvv$. All first-order fields (subscript \qmarks{1}) oscillate harmonically with the angular frequency $\omega$, whereas all second-order fields (subscript \qmarks{2}) are steady, being averaged in time over a full oscillation period $\frac{2\pi}{\omega}$.

The first-order acoustic pressure $p_1$ satisfies the Helmholtz equation in the bulk $\Omega$ and a boundary-layer boundary condition at $\pp\Omega$ expressed in terms of the inward normal derivative $\pp_\perp=-\nnn\cdot\nablabf$ and the outward-pointing normal vector $\nnn$~\cite{Bach2018},
 \bsub \eqlab{p1_gov} \begin{align}
 \Lapl p_1 + \kc^2 p_1&=0,\; \text{ inside } \Omega,\\
  \Big[\pp_\perp + \frac{\ii}{\ks} (\kc^2+\pp_\perp^2)\Big]p_1
  &=\frac{-\rhofl\omega^2}{1-\ii\Gamfl}\Big(\nnn+\frac{\ii}{\ks} \grad\Big) \scap \uuu_1^\mr{bdr},
  \nn \\
  &\quad  \text{ at the boundary } \pp\Omega.
 \end{align} \esub
Here, $\kc=\big(1+\frac12 \ii \Gamfl\big) \kO$ is the complex-valued compressional wave number having the real part $k_0=\frac{\omega}{\cfl}$, $\Gamfl=\frac12 \big(\frac43 +\frac{\etaBfl}{\etafl}\big) (k_0\delta)^2$ is the minute acoustic damping coefficient with $\Gamfl\ll1$, and $\ks=\frac{1+\ii}{\delta}$ is the shear wave number related to the viscous boundary layer of thin width $\delta=\sqrt{\frac{2\etafl}{\omega \rhofl}}$ with $k_0\delta \ll1$. From $p_1$, we obtain the acoustic velocity $\vvv_1$ and density $\rho_1$ outside the thin viscous boundary layer~\cite{Bach2018},
\begin{equation} \eqlab{v1}
 \vvv_1=\frac{-\ii(1-\ii\Gamfl) }{\omega\rhofl} \grad p_1, \qquad \rho_1=\kapfl p_1,
\end{equation}
with the isentropic compressibility $\kapfl = \frac{1}{\rhofl \cfl^2}$. The space-and-time-averaged acoustic energy density $\Eac$ in $\Omega$ of volume~$\calV_\Omega$~is,
 \begin{equation} \eqlab{Eac}
  \Eac = \int_\Omega \bigg[\frac14 \kapfl \big| p_1 \big|^2 + \frac14 \rhofl \big| \vvv_1 \big|^2 \bigg]
  \frac{\dd V}{\calV_\Omega}.
 \end{equation}
The second-order steady boundary-driven streaming velocity $\vvv_2$ outside the  viscous boundary layer is a Stokes flow with the slip velocity $\vvv_2^\mr{slip}$ at the boundary~\cite{Bach2018},
 \bsub \eqlab{v2_gov} \begin{align}
 &\zerovec = -\grad p_2 + \etafl \lap \vvv_2,\\
 &\text{with }\; 0 =\div\vvv_2  \text{ in } \Omega,\;\;
 \text{ and }\;\;
 \vvv_2 =\vvv_2^\mr{slip}  \text{ at } \pp\Omega.
 \end{align} \esub
For the slip velocity $\vvv_2^\mr{slip}$, we use expression~(55) of Ref.~\cite{Bach2018} for an oscillating, curved surface with a curvature radius much larger than the viscous boundary-layer width $\delta$.

The time-averaged forces acting on a suspended particle of radius $a$ and velocity $\vvv_\mr{pa}$ are the Stokes drag force $\FFFdrag$ and the acoustic radiation force $\FFFrad$~\cite{Settnes2012},
 \bsub \eqlab{particle_forces} \begin{align}
 \FFFdrag &= 6\pi\etafl a (\vvv_2-\vvv_\mr{pa}),\\
 \FFFrad   &= -\grad \Bigg[ \frac{4\pi a^3}{3} \bigg( \frac{f_0}{4} \kapfl |p_1|^2-\frac{3 f_1}{8} \rhofl |v_1|^2\bigg)\Bigg],
 \end{align} \esub
where $f_0$ and $f_1$ are the monopole and dipole scattering coefficients for the particle.  All parameter values are given in the Supplemental Material \footnote{See Supplemental Material at [URL] for a list of material parameters and a description of the shape optimization for a fluid domain embedded in an elastic solid.}.

\textit{Shape optimization for suppression of acoustic streaming.}---We consider straight microchannels placed along the $x$ axis with different $y$-$z$ cross sections, see \figref{rect_vs_shape}. To quantify the comparison between these channels, we revert to the classical results for a standing half-wave resonance $p_1= p_a \sin(k_0 y)$ in a rectangular cross section, for which $\Eac= \frac14 \kapfl p_a^2$, the slip velocity $\vvv_2^\mr{slip}=\frac{3\Eac }{2\rhofl\cfl}\sin(2 k_0 y)\eey$~\cite{LordRayleigh1884}, and the acoustic radiation force $\FFFrad=-4\pi a^3 k_0 \Phi\Eac \sin(2 k_0 y)  \eey$, with the acoustic contrast factor  $\Phi= \frac13 f_0 + \frac12 f_1$~\cite{Yosioka1955}. We then introduce the following characteristic scaling quantities based on the acoustic energy density $\Eac$~\eqnoref{Eac}: the acoustic pressure $p_1^0$, the streaming speed $\vscale$, and the radiation force $\Fscale$,
 \begin{equation} \eqlab{scaling}
 \pscale  = \sqrt{\frac{4\Eac}{\kapfl}}, \quad
 \vscale  = \frac{3\Eac}{2\rhofl\cfl} , \quad
 \Fscale  = 4\pi a^3 k_0 \Phi\Eac.
 \end{equation}
To optimize the shape for suppression of the acoustic streaming, we define a cost function $\Ccost$ that penalizes large streaming,
 \begin{equation} \eqlab{cost_function}
 \Ccost = \frac{1}{\vscale}\int_\Omega \abs{\vvv_2}\: \frac{\dd V}{\calV_\Omega}.
 \end{equation}
The suppression of the acoustic streaming is quantified by the suppression parameter $S_q$, the volumetric fraction in which the streaming speed $|\vvv_2|$ is smaller than the percentage $q$ of $\vscale$,
 \begin{equation} \eqlab{Sq}
 S_q = \int_\Omega \Theta\Big(\frac{q}{100}\vscale-\abs{\vvv_2}\Big)\: \frac{\dd V}{\calV_\Omega}.
 \end{equation}
Here, $\Theta(x)$ is Heaviside's step function being 0 for $x<0$ and 1 for $x>0$.

For a given cross-section shape, we evaluate the cost function $\Ccost$ by the following numerical two-step simulation in COMSOL Multiphysics~\cite{Comsol54}, see \eg\ Refs.~\cite{Muller2012, Bach2018, Skov2019}: (1) We compute $p_1$ from \eqref{p1_gov} in the idealized case of a prescribed displacement $d_0$ of the wall in the $y$ direction, $\uuu^\mr{bdr}_1 = d_0 \eey$. (2) We solve \eqref{v2_gov} for $\vvv_2$ with $\vvv_2^\mr{slip}$ calculated from $p_1$.

The cross-section shape is constrained to have width $W_0 = 380~\SImum$ and height $H_0 = 160~\SImum$, and to be symmetric in $y$ and $z$, see \figref{rect_vs_shape}(b). The upper right edge is represented by a cubic spline interpolation through $7$ points $(y_i,z_i)$, $i = 0,1,\ldots,6$, where the $y$ positions $y_i$ are fixed at $[0,\frac16,\frac36,\frac46,\frac56,\frac{9}{10},1] \frac{W_0}{2}$.  Furthermore, the $z$ positions of the end points are fixed at $z_0=\frac12 H_0$ and $z_6=\frac12 h_0$, where $h_0=10~\SImum$ is the height of the channel at the neck $y_6=\frac{1}{2}W_0$. The optimization algorithm minimizes the cost function
$\Ccost$~\eqnoref{cost_function} by varying the five free heights $z_1$-$z_5$ with the constraint $\frac12 h_0 \leq z_i \leq \frac{1}{2}H_0$. This optimization is implemented in Matlab~\cite{Matlab2019b} using the routine \texttt{fminsearchbnd}~\cite{fminsearchbnd} that calls COMSOL. It typically requires $\sim\!200$ iterations, each taking 5 seconds  on a workstation with a 3.5-GHz Intel Xeon CPU E5-1650 v2 dual-core processor and with a memory of 128~GB~RAM.

In \figref{rect_vs_shape}, simulation results are shown for the well-studied rectangular cross section~\cite{Muller2012} and compared to the results for the optimized spline cross section. For the optimized shape, the acoustic streaming is dramatically suppressed, whereas the radiation force is still present in the entire channel. Quantitatively, we obtain from \eqref{Sq} the streaming-suppression parameters $S_5=96\ \%$ and $S_1=63\ \%$ for the optimized shape, and $S_5=4\ \%$ and $S_1=0.6\ \%$ for the rectangle.

\begin{figure}[t]
\includegraphics[width=\columnwidth]{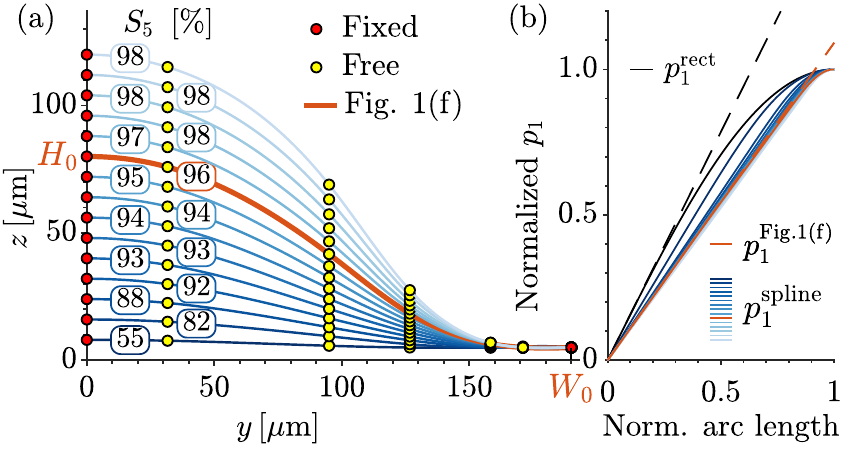}
\caption{\figlab{shape_family} (a) The optimized shapes, each defined by 2 fixed (red) and 5 free (yellow) points, obtained as in \figref{rect_vs_shape} but for different maximum height constraints ranging from $0.1H_0$ (dark blue) to $1.5H_0$ (light blue). The curves are labeled by the suppression parameter $S_5$, see \eqref{Sq}. The thick orange shape is the one shown in \figref{rect_vs_shape}(f). (b) The acoustic pressure $p_1$ versus arc length along the boundaries shown in  (a) using the same color scheme. The black curve shows the pressure obtained in the rectangular cross section, where $p_1^\mr{rect}\propto \sin(\frac{\pi y}{W})$, and the dashed lines are selected tangents.}
\end{figure}

In \figref{shape_family}(a), we show the family of optimized shapes obtained as above, but varying the maximum height as $H = [0.1,0.2,\dots,1.5]\:H_0$. In \figref{shape_family}(b), we plot $p_1$ along the upper boundary and note that it is approximately linear along a large part of the arc length for all the optimized shapes. This may be explained by inspecting the simplified expression for the slip velocity $\vvv_2^\mr{slip}$
adapted from Eq.~(61) in Ref.~\cite{Bach2018} 
to the 2D standing-wave resonance considered here,
 \begin{equation}\eqlab{vslip_simple}
 \vvv_{2\parallel}^\mr{slip} \approx -\frac{3}{8\omega } \grad_\parallel \big|\vvv_{1\parallel} \big|^2, \qquad v_{2\perp}^\mr{slip} \approx  0.
 \end{equation}
Clearly, because  $\vvv_{1\parallel} \propto \pargrad p_1$, the tangential slip velocity  $\vvv_{2\parallel}^\mr{slip}$ is small when $p_1$ is linear along the boundary. Remarkably, as seen in \figref{shape_family}(b), this linearity is maintained along nearly $90~\%$ of the optimized boundaries, but eventually, due to the no-slip boundary condition, the pressure gradient must tend to zero at the end-point $(\frac12 W_0, \frac12 h_0)$. The last 10~\% of the boundary therefore generates streaming, so by forming narrow necks there, the streaming becomes localized in a small region.

\begin{figure}[b]
\includegraphics[width=\columnwidth]{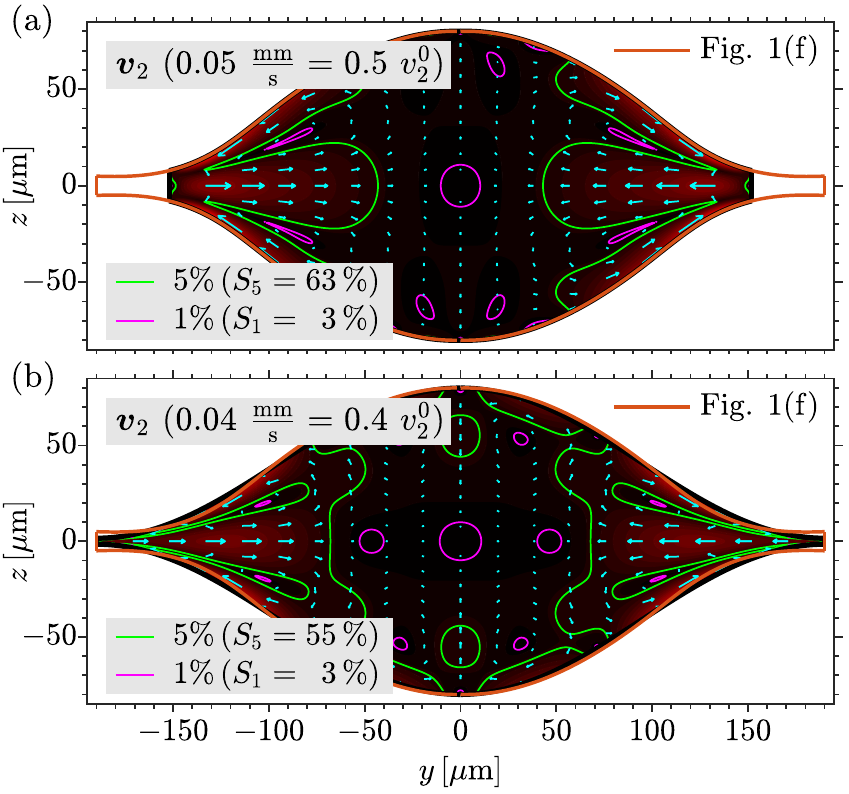}
\caption{\figlab{cut_cosine}
The acoustic streaming velocity $\vvv_2$ (cyan arrows), its magnitude $v_2$ from 0 (black) to 0.05~$\frac{\SImm}{\SIs}$ (light yellow), and the 5-\% (light green) and 1-\% (dark magenta) contour lines of $v_2$ as in \figref{rect_vs_shape}(f), but for two different
hard-wall shapes. (a) The optimized shape \figref{rect_vs_shape}(f) (orange dashed curve) with its necks cut off, and (b) an optimized cosine shape.}
\end{figure}

In \figref{cut_cosine}(a), we study the importance of the narrow necks of the shape in \figref{rect_vs_shape}(f) by cutting them off, leaving 90~\% of the width, $-\frac{9}{20} < \frac{y}{W_0} < \frac{9}{20}$. In this case, the streaming is still suppressed: with $S_5 = 63~\%$ and $S_1 = 3~\%$, it is worse compared to \figref{rect_vs_shape}(f) with the necks, where  $S_5 = 96~\%$ and $S_1 = 38~\%$, but much better than for the rectangle of \figref{rect_vs_shape}(e), where $S_5 = 4~\%$ and $S_1 = 1~\%$. As it might prove difficult in practice to fabricate the exact optimized shape, we study in \figref{cut_cosine}(b) a generic shape with a narrow neck and a wide bulk given by a cosine, $z(y) = \pm\frac{h}{2}\pm\big(\frac{H_0}{2}-\frac{h}{2}\big) \big[\frac12 + \frac12\cos(\frac{2\pi y}{W_0}\big) \big]$. Here, the neck height $h$ is the only free parameter. Using the cost function $\calC$ again, the optimal value is found to be $h=3.14~\SImum$ with a fair streaming suppression of $S_5 = 55~\%$ and $S_1 = 3~\%$. See Table~I of the Supplemental Material~\cite{Note1} for more simulation results on the channels with prescribed hard-wall motion.

\begin{figure}[b]
\includegraphics[width=\columnwidth]{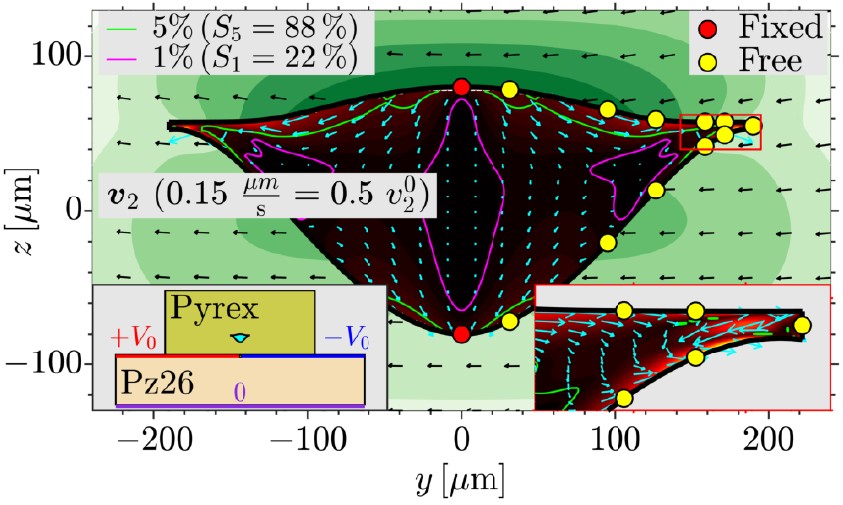}
\caption{\figlab{solid} The streaming field (black-to-red contour plot) in the fluid channel and the displacement field (light-to-dark-green contour plot) in the surrounding glass in a full-device simulation including a Pz26 transducer with split top and grounded bottom electrode (bottom-left inset).  The neck is shown in the lower-right inset. The optimized fluid channel shape is obtained by varying the $z$-coordinates of the 11 yellow points (the 2 red being fixed), see Supplemental Material for more details~\cite{Note1}.}
\end{figure}

\textit{Solids and transducers.}---More realistic models must include the elastic solid surrounding the fluid and the attached piezoelectric transducer. As shown in \figref{solid} (lower left inset), we embed the microchannel in a straight rectangular glass block (3~mm $\times$ 1.3~mm) mounted on a piezoelectric  transducer (5~mm $\times$ 1~mm) with a split top electrode for antisymmetric actuation by an AC voltage $\pm V_0 = \pm 1$~V as in the setup of Ref.~\cite{Moiseyenko2019}. As the up-down symmetry is broken, we now allow 11 free and 2 fixed points in the shape optimization, again using the cost function $\Ccost$. The resulting streaming field shows a fair suppression, $S_5 = 59~\%$ and $S_1 = 2~\%$, compared to the shape-optimized model with prescribed hard-wall motion  \figref{rect_vs_shape}(f), $S_5 = 96~\%$ and $S_1 = 38~\%$. See the Supplemental Material~\cite{Note1} for more details.

\textit{Particle focusing.}---In the conventional rectangular cross section, the minimum
radius $a_\mr{min}$ of particles that can be focused is estimated by equating  $\Frad^0$ and the drag force $6\pi\etafl a \vscale$~\cite{Barnkob2012a, Muller2012} from \eqref{scaling}, leading to $a_\mr{min}^\mr{rect}=\sqrt{\frac{9\etafl}{4\rhofl\omega\Phi}} =0.9~\SImum$ for polystyrene particles with $\Phi=0.16$ in water at $f=1.95~\SIMHz$. In the optimized shape, the streaming is suppressed by 95~\% at $f=3.02$~MHz leading to a substantial six-fold reduction of $a_\mr{min}$ to $a_\mr{min}^\mr{opt} \approx 0.15~\SImum$.

\textit{Conclusion.}---By exploiting effective boundary conditions~\cite{Bach2018}, we have implemented an optimization algorithm that computes the shape of an acoustic cavity, which at resonance has the remarkable property that the acoustic streaming is dramatically suppressed relative to the conventional rectangular cavity. Notably, the acoustic pressure amplitude and the acoustic radiation force acting on suspended particles are not suppressed, and therefore, the optimized cavity shape is particularly ideal for applications within controlled handling of nm-sized particles in acoustophoresis.

We have demonstrated how shape optimization can be used to gain insight in fundamental acoustofluidics, in particular how to suppress the ubiquitous acoustic streaming by ensuring a linear acoustic pressure profile along the wall; and how such an insight can be used for practical applications. By applying other optimization
methods, say topology optimization \cite{Olesen2006}, or other cost functions, such as one based on acoustophoretic force fields, our method may be extended to other fundamental studies within nonlinear acoustics.

%
%


%

\end{document}